\documentclass{PoS}

\def\sl#1{\rlap{\hbox{$\mskip 1 mu /$}}#1}
\def\Det{{\rm Det}}
\def\tr{{\rm tr}}
\def\Tr{{\rm Tr}}
\def\exp{{\rm exp}}
\def\ie{{\it i.e.}}
\def\leqx{\,\raisebox{-1.0ex}{$\stackrel{\textstyle <}{\sim}$}\,}
\def\psibar{{\overline{\psi}}}

\title{QCD with rooted staggered fermions}

\ShortTitle{Rooted staggered fermions}

\author{{Maarten Golterman}
\\
        Department of Physics and Astronomy, San Francisco State University,\\
        San Francisco, CA 94132, USA\\
        E-mail: \email{maarten@stars.sfsu.edu}}

\abstract{In this talk, I will give an overview of the theoretical status of staggered Lattice QCD with the
``fourth-root trick.''   In this regularization of QCD, a separate staggered quark field is used
for each physical flavor, and the inherent four-fold multiplicity that comes with the
use of staggered fermions is removed by taking the fourth root of the staggered
determinant for each flavor.  At nonzero lattice spacing, the resulting theory is nonlocal and
not unitary, but there are now strong arguments that this disease is cured in the
continuum limit.  In addition, the approach to the continuum limit can be
understood in detail in the framework of effective field theories such as 
staggered chiral perturbation theory.}

\FullConference{8th Conference Quark Confinement and the Hadron Spectrum \\
		 September 1-6 2008\\
		 Mainz, Germany}

\begin{document}

\section{Introduction}

In the last few years, it has become possible to compute many hadronic quantities of
phenomenological interest using Lattice QCD; for an overview of recent results, see
the talk by Kronfeld at this conference \cite{Kronfeld}.  Many of these results have been
obtained using gauge configurations that include the effects of three light dynamical
quarks, in which a (highly improved) staggered Dirac operator is used to discretize 
the quark action.  Staggered fermions are attractive because of the relatively
low expense required for reaching very light quark masses at very small lattice spacings.
For most of these results, the claim is that all errors, statistical and
systematic, are under control.  However, as I will describe in more detail below, in order
to remove a spurious four-fold redundancy inherent to staggered fermions, the fourth root of the
fermion determinant for each physical flavor is taken inside the integral over the gauge
field.  This raises the critical question whether this method constitutes a valid
regulator for QCD.  In this talk, I will describe the problem in some detail, and then discuss the, in my view interesting and important, 
progress that has been
made in answering this question.  This talk is meant to give an overview, rather than a 
complete review of all work in this direction, as 
I do not have enough space to be complete.
For other recent reviews, see Refs.~\cite{Sharpe,BGS06,AK,SD}, which contain many
more references to other relevant work.

Let me first very briefly recount the origin of the four-fold redundancy.  A naive nearest-neighbor
discretization of the free, massless Dirac operator, $S^{-1}(p)=i\sl{p}$, leads to
an inverse lattice propagator of the form ($a$ is the lattice spacing)
\begin{equation}
\label{lattprop}
S^{-1}(p)=\sum_\mu\frac{i}{a}\;\gamma_\mu\sin{(ap_\mu)}\ .
\end{equation}
In addition to the expected zero at $p=0$, $S^{-1}(p)$ has fifteen other zeros with at least
one component of $p$ equal to $\pi/a$ on the Brillouin
zone, from which it follows that this lattice fermion describes sixteen massless fermions
in the continuum limit.  This is an example of the well-known species doubling problem.  There is 
a deep reason for the occurrence of these doublers in terms of the axial anomaly:
a regulated theory with exact chiral symmetry has to produce an anomaly-free
representation in the continuum limit \cite{KSNN}. 

Staggered fermions \cite{KoSu} reduce this multiplicity by four.  They are constructed
from naive lattice fermions by dropping the Dirac index, and replacing the $\gamma$-matrices
by judiciously chosen, $x$-dependent phases.  This reduces the sixteen-fold doubling to a four-fold doubling.
In other words, each staggered fermion describes four degenerate relativistic flavors in the continuum
limit, which we will henceforth refer to as the four ``tastes'' of each staggered fermion.  The emergence
of this continuum limit, which carries over to the interacting case, is a consequence of lattice symmetries and dimensional analysis:
lattice symmetries guarantee that a continuum limit with SO(3,1) Lorentz and
SU(4)$_L\times$SU(4)$_R$ chiral ``taste'' symmetry is obtained without any tuning of the
action \cite{GoSm}.  A particularly important lattice symmetry is the U(1) transformation that rotates the
fermion fields on even and odd lattice sites with opposite phases (``U(1)$_\epsilon$ symmetry'') \cite{KaSm}, which is an exact axial symmetry, broken by a (single-site) mass term.

In practice, one uses the following method for simulating QCD with three light flavors.
A separate staggered field is introduced for each physical flavor, with single-site
mass terms for each, with masses $m_u$, $m_d$ and $m_s$.  Each of these
flavors thus comes in four tastes, and the theory would thus contain four up, four down and four
strange quarks in the continuum limit, with a U(4)$_u\times$U(4)$_d\times$U(4)$_s$ taste symmetry.
To eliminate this unphysical multiplicity, the fourth root of each staggered determinant
is taken, motivated by the observation that, in the continuum limit, the staggered determinant
should factorize as \cite{MPR}
\begin{equation}
\label{factorize}
\Det(D_{stag})\sim\Det^4(D_{cont})\ .
\end{equation}
Since for all $m\ne 0$, $\Det(D_{stag})>0$ (this determinant depends only on $|m|$ because
of $U(1)_\epsilon$ symmetry), it is clear that the positive fourth root should be taken, and that
the resulting quark mass $m_q\propto |m|$.\footnote{Quark mass renormalization is
multiplicative \cite{GoSm}.}  This, then, constitutes a regularization of three-flavor QCD with all quark masses positive.\footnote{For a discussion of negative quark
mass in the context of a two-dimensional one-flavor theory, see Ref.~\cite{DH}.}  The topic of this talk is
the validity of this regularization.

\section{The problem}

First, let us consider a continuum theory with exact U(4) taste symmetry,
\ie, with four fermions with equal positive quark masses.  In this case, 
one can take the fourth root of the fermion determinant, and doing so 
reduces the partition function to that a of theory with one flavor,
$N_f = 1$. In this rooted theory, one still has access to correlation 
functions with all four tastes on the external lines, {\it e.g.}, correlation 
functions of the fifteen pions of the four-taste theory. It is thus 
interesting to ask in exactly what sense taking fourth root reduces 
the number of pions from fifteen to none.

The key observation is that, since rooting reduces the number
of sea quarks from four to one, the correct number for the $N_f=1$ 
theory, it is possible to construct consistent projections into
the physical, unitary one-flavor theory \cite{BGSS}.  
I will illustrate this with an example in the meson sector.

For this, it is useful to describe rooting in terms of the replica rule: if we take
$n_r$ copies of a U(4)-taste fermion field, so that the four-taste 
fermion determinant appears raised to the
$n_r$-th power, then continuing $n_r\to 1/4$ corrresponds
to taking the fourth root.  Armed with this tool, let us consider, for example, the two-pion intermediate states in the
taste-singlet scalar two-point function in chiral perturbation theory (ChPT) \cite{CB,BDFP}.  
These two-pion states produce a cut starting
at $2m_\pi$, and in the theory with $n_r$ replicas, the ``strength'' of this cut is
$16n_r^2-1$, because that is the number of pions in the theory with $n_r$ replicas, which has SU$_L$(4$n_r$)$\times$SU$_R$(4$n_r$) chiral 
taste-replica symmetry for any positive integer $n_r$.  If we now continue $n_r\to 1/4$ (see also
Sec.~5 below),
we see that this factor vanishes, and the two-pion cut
disappears, as it should; this follows from taste symmetry.  Our example
demonstrates how the theory is unitary for $n_r=N_f/4$ for any positive integer $N_f$ even if $N_f<4$,
despite the presence of ``too many'' pions in the rooted theory.

On the lattice, taste symmetry is broken to a much
smaller, discrete group, and the argument above no longer holds.  This
raises three questions that we will consider in the rest of this talk:
\begin{itemize}
\item[1)]  Is rooted staggered QCD a regularization like any other, or not?  The answer is
no, the theory is nonlocal and nonunitary at $a\ne 0$.
\item[2)]  Can the continuum limit be taken, and is it in the correct universality class?
Here the answer is most likely yes, and we will briefly review the renormalization-group
(RG) based argument supporting this claim \cite{YS}.
\item[3)]  But this is not the end of the story.  We work at $a\ne 0$, and scaling violations,
while small, are still significant.  Hence, in actual computations, the diseases are
present, and one needs effective-field theory (EFT) techniques to parameterize the
nonlocal effects (in addition to the need to control continuum and chiral extrapolations).
The question is whether such an EFT framework exists.  The claim is that it
is provided by ``Staggered ChPT plus the replica rule,'' or
rSChPT for short \cite{SChPT}.  Two related, but different derivations have been given:  
The first derivation is entirely within the ChPT framework, and starts with the observation that rooting works trivially for a theory
with {\it four} degenerate staggered fermions, \ie, an $N_f=4$ theory.  One then moves to
the nondegenerate case by expanding around the degenerate case.  This allows one, under
certain assumptions, to decouple one or more of the fermions, thus arriving at the cases
$N_f=$1, 2, or 3 \cite{CB}.  The other is a direct derivation from the RG
framework of Ref.~\cite{YS} which we will describe below.
\end{itemize}

\section{Nonlocality and nonunitarity from taste-symmetry breaking.}

Taste symmetry is broken on the lattice, and we may thus split the staggered Dirac
operator into two parts, $D_{stag}=D\otimes{\bf 1}_4+a\Delta$, with ${\bf 1}_4$ the unit
matrix in taste space, and $\Delta$ the taste-breaking part (with $\tr_{taste}(\Delta)=0$).
The taste-breaking part vanishes linearly with $a$ in the classical continuum
limit, which is why I factored out the explicit factor $a$.
It follows that
\begin{equation}
\label{effaction}
\log\Det(D_{stag})=4\log\Det(D)+\log\Det\left(1+D^{-1}a\Delta\right)\ .
\end{equation}
While both $D$ and $a\Delta$ are local, clearly $D^{-1}a\Delta$ is not!  This means that 
taste breaking, while local at the level of the action, has nonlocal consequences for the
physics.   Indeed, the second term on the right breaks taste symmetry, and lifts the degeneracy 
of the fifteen pions of the theory defined by $D\otimes{\bf 1}_4$.
The pion spectra of the $D_{stag}$ and $D\otimes{\bf 1}_4$ theories do not match.
{}From this observation, it is easy to prove that the rooted theory is 
nonlocal at $a\ne 0$ \cite{BGS}.

Lowest-order SChPT gives the pion masses of the staggered theory as
\begin{equation}
\label{pionmasses}
(m_\pi^A)^2=Bm_{quark}+c^A a^2\Lambda_{QCD}^4\ ,
\end{equation}
in which the index $A$ labels the different pions, which fall into irreps of the exact remnant of
taste symmetry on the lattice \cite{MG}, with a different value of $c^A$ for each irrep.\footnote{
There is only one exact Goldstone boson, for which $c^A=0$ in Eq.~(\ref{pionmasses}), as a consequence of $U(1)_\epsilon$
symmetry.}   
The pion-mass behavior predicted by Eq.~(\ref{pionmasses}) is clearly seen in numerical simulations \cite{MILC}.  
Figure~1 shows the nondegeneracy of the various different pions, with the various labels corresponding
to values of the index $A$ (for details, we refer to Ref.~\cite{MILC}), and Fig.~2 shows how the
taste splittings scale with the lattice spacing ($r_1$ is a quantity used to set the scale).

\begin{figure}[t]
\begin{center}
\parbox[t]{0.45\linewidth}{
\centerline{\includegraphics[width=0.94\linewidth]{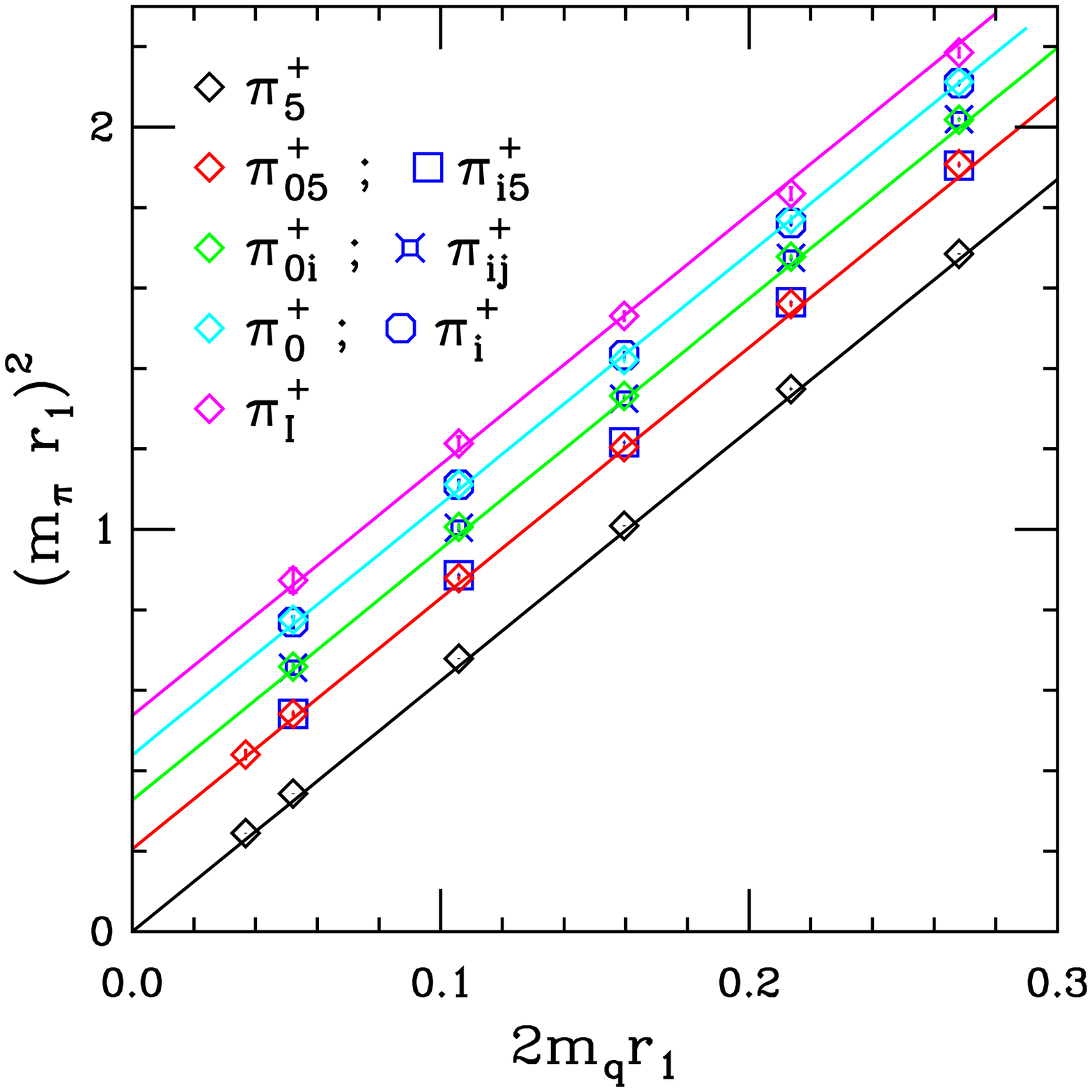}}
\caption{Squared pion masses as a function of the light quark mass for $a=0.12$~fm
(from Ref.~\cite{MILC}).}
\label{FULL}
}
\hspace{0.05\linewidth}
\parbox[t]{0.45\linewidth}{
\centerline{\includegraphics[width=\linewidth]{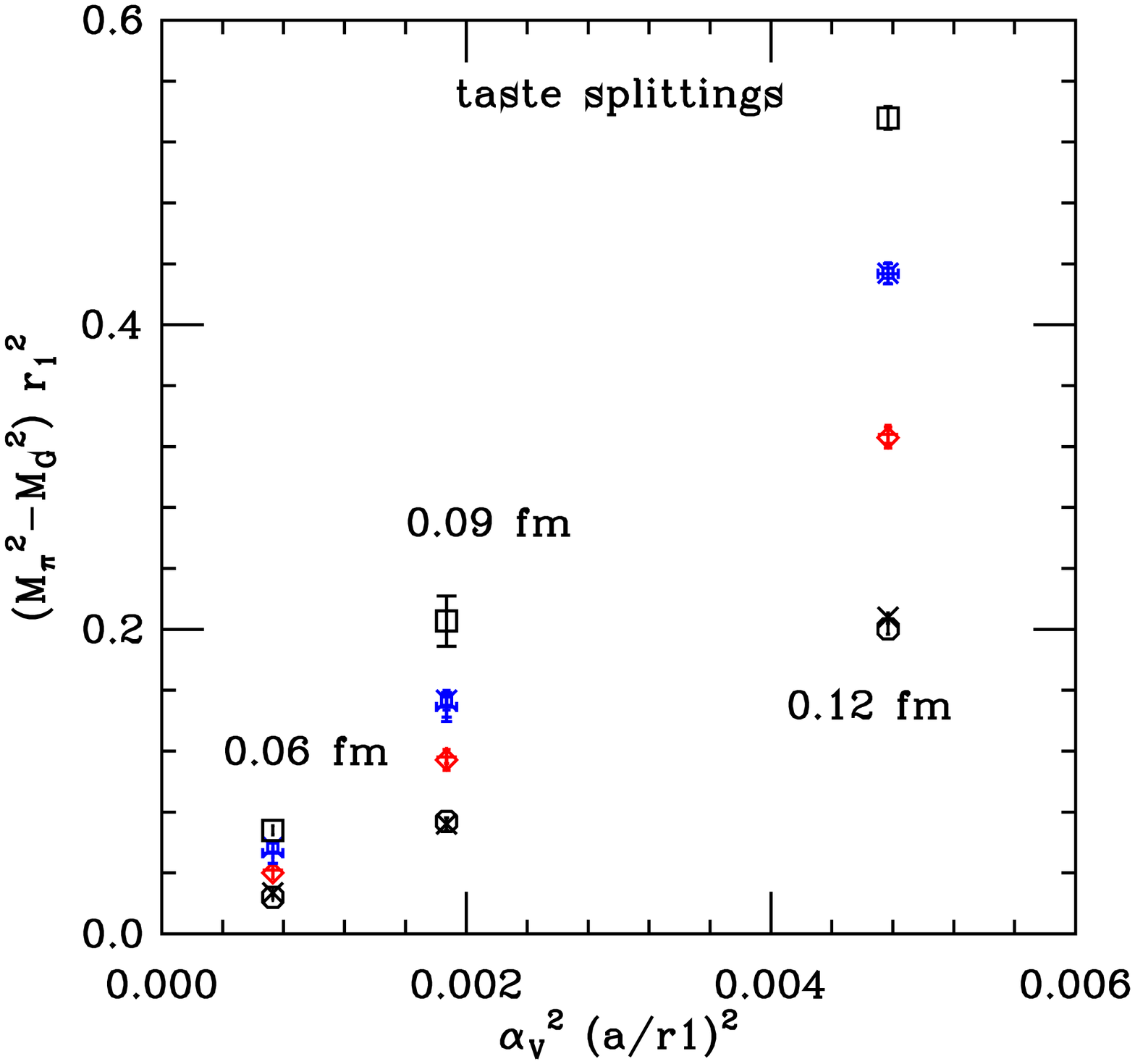}}
\caption{Taste splitting among the pions as a function of lattice spacing
(from Ref.~\cite{MILC}). $M_G$ is the mass of the exact Goldstone boson.}  
\label{TASTE}
}
\end{center}
\end{figure}

This pattern implies that the $N_f=1$ ``pion counting'' argument given at the beginning
of the previous section for the continuum rooted theory is violated on the lattice.  Exact 
cancellations that occurred because of full SU(4$n_r$) taste symmetry no longer occur \cite{BDFP,CB},
and an $O(a^2)$
two-pion cut survives in the taste-singlet two-point function for $n_r=1/4$.  
This example demonstrates that
indeed rooted staggered fermions are not unitary at $a\ne 0$, and that this violation of
unitarity occurs at physical scales, exhibiting the nonlocality of the theory.

While this discussion confirms the sickness of the rooted theory at $a\ne 0$, it teaches us
several important things.  First, if the taste-breaking operator $a\Delta$ is irrelevant 
(in the sense of the RG),  the nonlocal and nonunitary
behavior disappears in the continuum limit.  In the unrooted theory, $a\Delta$ is indeed
irrelevant, and taste symmetry is restored in the continuum limit.  However, this is not obvious in the rooted theory, since it requires
the extension of RG techniques to the nonlocal theory at $a\ne 0$.  We will investigate this
in the next section.  

Another important observation is that pion masses are governed by two different IR
scales: the physical quark mass $m$, and the unphysical taste splitting $(a\Lambda_{QCD}^2)^2$
that leads to unitarity violations in the $a\ne 0$ rooted theory.   It is thus clear that (a) the limit
$m\to 0$ at $a\ne 0$ is unphysical \cite{CBlim}, and (b) that the limit $a\to 0$ has to be taken before
continuation to Minkowski space.

\section{Continuum limit: a renormalization-group framework.}

It is natural to study the approach of the continuum limit in an RG framework.  First,
it gives us a tool for a precise definition of the continuum limit, making it possible to
define what is meant by the intuitive factorization of the staggered determinant, Eq.~(\ref{factorize}).
Second, in the unrooted theory, one expects that low-lying (IR) eigenvalues form 
taste quadruplets when $a$ becomes small, while this won't happen for the UV (cutoff)
eigenvalues.  Here RG blocking helps: it gets rid of the UV eigenvalues.\footnote{For
numerical investigations of staggered eigenvalues, see Ref.~\cite{EV}.}

A simple hypercubic blocking scheme can be set up \cite{YS} in which one takes the
``coarse'' lattice spacing $a_c\ll 1/\Lambda_{QCD}$ arbitrarily small but fixed, and the fine lattice spacing
$a_f$ to zero: after $n$ blocking steps, the relation between the two lattice spacings is given by
$a_c=2^na_f$.  The usual universality arguments imply that this will lead to the expected continuum
limit for unrooted staggered fermions because they are local, and this is what we will assume in the rest of this talk.  
In contrast, no direct RG blocking can be defined
for the rooted theory, since the rooted theory is not formulated in terms of a path integral:
the fourth root is taken inside the integral over gauge fields after the fermionic integral
has been performed.  However, it is possible to construct a bridge between the unrooted
and rooted theory at each blocking step: these ``reweighted'' theories will be constructed to have
the same $a_f\to 0$ theory as the staggered theory, 
but they also will have exact taste symmetry \cite{YS}.

This works as follows.  After each blocking step, we split the (blocked) staggered
Dirac operator $D_{stag,n}=D_n\otimes{\bf 1}_4+a_f\Delta_n$, with $a_f\Delta_n$ the
taste-breaking part, just as before Eq.~(\ref{effaction}).\footnote{One can ``postpone'' the
integration over the gauge fields on the blocked lattices, so that the fermion integration
remains gaussian at each step \cite{YS}.}  The {\it unrooted}
staggered theory defined by $D_{stag,n}$ and the taste-invariant theory defined by
$D_n\otimes{\bf 1}_4$ have the same continuum limit, because $a_f\Delta_n$ will scale as
expected in this case.  The theory defined by $D_n\otimes{\bf 1}_4$
has exact taste symmetry, and is local on the $n$-th lattice.  This theory can thus be rooted,
and one obtains a {\it local} one-taste theory with partition function
\begin{equation}
\label{reweigh}
Z^{reweigh}=\int d\mu_{gauge}\;\Det(D_n)\ .
\end{equation}
The claim is now that for $n\to\infty$ (at fixed $a_c$) this local theory coincides with the 
nonlocal theory
\begin{equation}
\label{root}
Z^{root}=\int d\mu_{gauge}\;\Det^{1/4}(D_{stag,n})\ .
\end{equation}
Indeed, if $||a_f\Delta_n||\leqx a_f/a_c$, one has that
\begin{equation}
\label{expansion}
\Det^{1/4}\left(D_n\otimes{\bf 1}_4+a_f\Delta_n\right)
=\Det(D_n)\;\exp\left[\frac{1}{4}\tr\log\left({\bf 1}_4+D_n^{-1}a_f\Delta_n\right)\right]
=\Det(D_n)\left(1+O\left(\frac{a_f}{a_c^2m}\right)\right)\ .
\end{equation}
For this to work, it is necessary that $a_f\Delta_n$ scales like $a_f/a_c$ on an ensemble.  
We need this scaling between $1/a_f$ and $1/a_c\ll 1/a_f$, and, since $1/a_c\gg\Lambda_{QCD}$, this
scaling should be calculable in perturbation theory.  The conjecture here is that one can
rely on this to work, because it concerns the scaling of a local operator ($\Delta_n$), in a 
renormalizable theory.\footnote{There is no space for a detailed discussion of this
point, for which we refer to Ref.~\cite{YS}.}

We may rephrase the RG argument as follows \cite{BGS06}:
\begin{itemize}
\item[-]  The starting point is that $a_f\Delta_n$ scales like $a_f$ in the unrooted
staggered theory, because this theory is local.  It has to, if the expected continuum limit for this theory exists.
\item[-]  Therefore, $a_f\Delta_n$ scales like $a_f$ in the four-taste reweighted theory
defined by $D_n\otimes{\bf 1}_4$, which is U(4) taste invariant and local on the $n$-th lattice.
\item[-]   One then expects that $a_f\Delta_n$ scales like $a_f$ in the {\it one-taste} reweighted
theory; because of the exact taste symmetry of reweighted theories, the one-taste reweighted theory 
is still local.
\item[-]  Finally, one may reconstruct the rooted staggered theory from the one-taste
reweighted theory, using the expansion (\ref{expansion}).  
\end{itemize}
We end this section with the comment that, clearly, a necessary condition for all this to work is that rooting works 
in perturbation theory.  Indeed, it does: in the theory with $N_f$ flavors and $n_r$ replicas,
the total number of quarks on any closed loop is equal to $4N_fn_r$, which, for $n_r=1/4$
(which corresponds to taking the fourth root of each of the $N_f$ staggered determinants)
is precisely equal to $N_f$ \cite{BGPQ,Sharpe}.  It follows that indeed the rooted theory is
(perturbatively) renormalizable, and thus standard power counting, according to which
$a_f\Delta_n$ is irrelevant, applies.

\section{Staggered ChPT from the RG approach.}

After reviewing the RG-based argument for the validity of rooting in the continuum limit, 
we now use this framework to derive the
existence of an EFT framework for the rooted staggered theory at nonzero $a$, thus addressing
the concern expressed in the third question of Sec.~2 \cite{BGSEFT}.  EFTs such as the
Symanzik effective theory (SET) \cite{Sym} and ChPT account for lattice artifacts through a systematic
expansion in $a\Lambda_{QCD}$.  An example may illustrate this as follows.  The taste breaking
at $a\ne 0$ leads to taste-breaking four-fermion operators in the effective continuum theory,
much like ``new physics'' at a higher scale leads (for example) to effective four-fermion operators to 
be added to the Standard Model action.  The ``new physics'' here is the taste (and
rotational) symmetry breaking in the underlying lattice theory.
For instance, the SET for the staggered theory contains an
operator of the form
\begin{equation}
\label{ff}
a^2(\psibar_R\xi_\nu\xi_5\psi_L)(\psibar_R\xi_\nu\xi_5\psi_L)+{\rm h.c.}
\ \ \ \to\ \ \ a^2\tr[\xi_\nu\xi_5\Sigma\xi_\nu\xi_5\Sigma]+{\rm h.c.}\ ,
\end{equation}
in which the $\xi_\nu$ are a set of $4\times 4$ $\gamma$-matrices acting in taste space.
On the right-hand side, I gave the translation of this four-fermion operator into ChPT, in terms
of the nonlinear pion field $\Sigma$.  Of course, all such operators, and their translation into
ChPT, have to be systematically classified \cite{SChPT}.

The key assumption on which the existence of EFTs is founded is that the underlying
theory, in this case, the lattice theory, is local.  Since the rooted staggered theory is not local,
the construction of EFTs like the SET and ChPT along the lines described above is not
automatic.  The question is thus whether the construction of a SET and staggered ChPT can
be extended to rooted staggered QCD.

The replica rule of Sec.~2 gives us an intuitive idea of what to do, but there is a catch.  
One starts with a theory
with $n_r$ staggered fermions, with $n_r$ a positive integer.  One constructs the desired
EFT, and simply continues $n_r\to N_f/4$ in this EFT,\footnote{I will restrict the discussion to
$N_f<4$ degenerate flavors; the generalization to nondegenerate masses is obvious.}  since this is precisely how one obtains
the theory with $N_f$ flavors from staggered QCD through rooting (if $N_f$ itself is not a
multiple of four).  In other words, one continues the EFT from integer values of $n_r$,
where the underlying theory is local, to quarter-integer values.  This should work for the explicit
dependence on $n_r$ that comes from calculating diagrams with loops in the EFT.  The catch
is, however, that the EFT depends on $n_r$ not only through loops, but also through the
coupling constants that multiply the operators which build up the EFT.    As long as $n_r$ is
integer, these coupling constants are uniquely determined by the underlying local theory.
But for quarter-integer values, they have to be obtained by continuation, and a
unique continuation off the positive integers does not exist.  Moreover, it might happen that 
the continuation will encounter a singularity precisely at $n_r=N_f/4$.  All this implies that we 
need more information about the dependence on $n_r$ of the correlation functions of the
underlying lattice theory.

This is where the RG framework of the previous section comes in.  First, take $n_r$ a positive
integer, and carry out $n$ RG blocking steps.  The resulting theory has a partition function
\begin{equation}
\label{replicaZ}
Z(n_r)=\int d\mu_{gauge}\;\Det^{n_r}(D_{stag,n})\ .
\end{equation}
Now, we generalize this theory by replacing (recall, $D_{stag,n}=D_n\otimes{\bf 1}+a_f\Delta_n$)
\cite{BGSEFT}
\begin{equation}
\label{generalize}
\Det^{n_r}(D_{stag,n})\to\Det^{n_s}(D_n)\;
\frac{\Det^{n_r}(D_n\otimes{\bf 1}+t a_f\Delta_n)}{\Det^{n_r}(D_n\otimes{\bf 1})}\ .
\end{equation}
Here $n_s$ is the desired number of physical flavors (with a given quark mass), and we thus need
$n_s/4$ staggered quarks.  At this point, however, we still keep $n_r$ integer, and not
necessarily equal to $n_s/4$.  Note the new ``interpolating'' parameter $t$.  We make the
following observations:
\begin{itemize}
\item[-]  For $n_s=4n_r$ and $t=1$ this is the staggered theory with $n_r$ replicas, hence
the right-hand side of Eq.~(\ref{generalize}) indeed generalizes the theory (\ref{replicaZ});
\item[-]  For $t=0$ this is the (local!) reweighted, taste-invariant theory with $n_s$ taste-singlet fermions;
\item[-]  For $n_s\ne 4n_r$ (and $t\ne 0$), this is a partially quenched theory \cite{BGPQ},
in which the determinant in the denominator is obtained from a path integral over
``ghost'' quarks with opposite (\ie, bosonic) statistics.
\end{itemize}
As long as $n_r$ and $n_s$ are positive integers, and for any $t$, this defines a local,
but partially quenched theory.   Our key assumption will be that for such theories
EFTs like the SET and ChPT exist.\footnote{This has become ``standard lore'' in 
lattice gauge theory, and there is now rather extensive numerical evidence supporting
the validity of this assumption.  This point has also been emphasized in Refs.~\cite{CB,Sharpe}.}

What this setup buys us is the following.  Expanding the determinant ratio in Eq.~(\ref{generalize}) using
\begin{equation}
\label{texpansion}
\Det^{n_s}(D_n)\;\frac{\Det^{n_r}(D_n\otimes{\bf 1}+t a_f\Delta_n)}{\Det^{n_r}(D_n\otimes{\bf 1})}
=\Det^{n_s}(D_n)\;\exp\left[n_r\;\Tr\log\left({\bf 1}+t(D_n^{-1}\otimes{\bf 1})a_f\Delta_n\right)\right]\ ,
\end{equation}
one sees that, in this expansion, the power of $n_r$ is smaller than the power of $t$, which, in
turn, is smaller than or equal to the power of $a_f$ to which we expand.  It follows that all correlation
functions of the theory, when expanded to some fixed order in $a_f$, are {\it polynomial}
in $n_r$!  Since this is true in the underlying lattice theory, it has to be true in any EFT
representing this theory, and we may thus continue $n_r$ to $n_s/4$ in the EFT.    In the end,
we may also set $t=1$, thus arriving at the EFT for the original staggered theory with $n_r$
replicas, but now for any $n_r=n_s/4$, with $n_s$ a positive integer.  The correctness of
rSChPT thus follows directly from the RG argument that supports the conjecture that
rooted staggered fermions constitute a regularization of QCD in the correct universality class.

Note that the argument sketched above does not imply that we have to actually perform the
continuation off integer values of $n_r$ explicitly.  The point is that our argument proves that
the values of the coupling constants of the EFT are uniquely determined by the underlying
lattice theory.  It follows that the desired values (those at $n_r=n_s/4$ and $t=1$)
can then be determined by fits to the numerically
computed correlation functions of the rooted theory itself.

One may ask why this approach does not imply that the theory may be defined for {\it any}
(real) value of $n_r$.  The key point here is that, as should be clear from the continuum
example at the start of Sec.~2, only for $n_r=n_s/4$ with positive integer $n_s$ the continuum
limit corresponds to a unitary theory.\footnote{Perturbative renormalizability holds indeed for
any $n_r$.}  A corollary is that rSChPT should reproduce the
sicknesses of the rooted staggered theory at $a\ne 0$, and indeed it does.  A
nontrivial test of this was performed in Ref.~\cite{BDFP}, in which the $a_0$ and $f_0$
two-point functions were fitted to rSChPT.  The values of low-energy constants found with this fit are
in good agreement with those fitted from pion and kaon masses and decay constants.

\section{Conclusions}

While I have only been able to give a very schematic overview of (some of) the arguments,
I conclude that, while at nonzero lattice spacing rooted staggered QCD is nonunitary,
it is very likely to have the correct continuum limit.  The RG-based arguments, in particular,
tie the validity of the rooted theory very strongly to the --- uncontested --- validity of the local,
unrooted theory.  

In addition, I have shown how one can derive EFTs, such as rSChPT, which are valid
at $a\ne 0$.  This is necessary both because of the fact that scaling
violations, while small, are not negligible at present, and also in order to test our understanding
of the nonphysical effects of rooting numerically.  The validity of rSChPT makes it 
possible to do numerical computations with pion masses down to $m_\pi^2\sim a^2\Lambda_{QCD}^4$,
which is crucial, with present resources, for reliable extrapolations to the physical values of the up and down
quark masses.  I emphasize that, since rSChPT follows directly from the RG argument
for rooted staggered QCD, fits of numerical data using rSChPT constitute direct tests of this
argument for the 
validity of rooting.  An interesting test in this respect is an rSChPT fit in which $n_r$ was
kept as a free parameter in the fit, yielding $n_r=0.28(4)$ \cite{MILCnr}.

In conclusion, there is now very good theoretical and numerical evidence that using the
fourth-root trick works, despite the fact that for $a\ne 0$ the theory is sick.  There is at present
no valid argument that the fourth root trick fails (see the Appendix for an additional comment).

\section*{Acknowledgments}

First, I would like to thank the organizers of Confinement{\it 8} for the opportunity to
present the current status of this important problem in Lattice QCD, and for an interesting and  well-organized conference.
I also thank Claude Bernard, Andreas Kronfeld, Yigal Shamir and Steve Sharpe for many useful discussions and comments.  This work is supported in part by the
US Dept. of Energy.

\section*{Appendix}

The only published arguments against rooting are those of Creutz (Ref.~\cite{MC} and refs.
therein).  I will not
revisit the discussion of these arguments here again, since they have been proven
incorrect \cite{BGSS,AK} in all their incarnations.   Indeed, none
of our detailed arguments refuting his claims have been addressed
by Creutz; in Ref.~\cite{MC} they are simply ignored.
I emphasize that refuting Creutz's arguments
by itself does not prove rooting to be correct, and I have reviewed the current
status of the evidence for the validity of rooting in this talk.

\end{document}